\documentclass[aps,pre,reprint,superscriptaddress,longbibliography]{revtex4-1}

\usepackage{graphicx}
\usepackage{times}
\usepackage{epstopdf}

\usepackage{amssymb}	
\usepackage{amsfonts}
\usepackage{amsmath}
\usepackage{mathrsfs}
\usepackage{mathtools}
\usepackage{times}
\usepackage{xcolor}
\usepackage{adjustbox}

\newcommand{\thetas}{\boldsymbol{\theta}}
\newcommand{\zetas}{\boldsymbol{\zeta}}
\newcommand{\etas}{\boldsymbol{\eta}}
\newcommand{\ps}{\boldsymbol{p}}
\newcommand{\qs}{\boldsymbol{q}}

\newcommand{\hatqs}{\boldsymbol{\hat{q}}}

\newcommand{\ac}{{a}}

\begin{document}


\title{Node metadata can produce predictability transitions in network inference problems }



\author{Oscar Fajardo-Fontiveros}
\email{oscar.fajardo@urv.cat}
\affiliation{Department of Chemical Engineering, Universitat Rovira i Virgili, 43007 Tarragona, Catalonia}
\author{Marta Sales-Pardo}
\email{marta.sales@urv.cat}
\affiliation{Department of Chemical Engineering, Universitat Rovira i Virgili, 43007 Tarragona, Catalonia}
\thanks{Corresponding author}
\author{Roger Guimer\`a}
\email{roger.guimera@urv.cat}
\thanks{Corresponding author}
\affiliation{ICREA, 08010 Barcelona, Catalonia}
\affiliation{Department of Chemical Engineering, Universitat Rovira i Virgili, 43007 Tarragona, Catalonia}


\date{\today}

\begin{abstract}
    Network inference is the process of learning the properties of complex networks from data. Besides using information about known links in the network, node attributes and other forms of network metadata can help to solve network inference problems. Indeed, several approaches have been proposed to introduce metadata into probabilistic network models and to use them to make better inferences. However, we know little about the effect of such metadata in the inference process. Here, we investigate this issue. We find that, rather than affecting inference gradually, adding metadata causes abrupt transitions in the inference process and in our ability to make accurate predictions, from a situation in which metadata does not play any role to a situation in which metadata completely dominates the inference process. When network data and metadata are partly correlated, metadata optimally contributes to the inference process at the transition between  data-dominated and  metadata-dominated regimes.
\end{abstract}


\maketitle


Many systems can be represented as networks, with nodes representing units (for example, people in a social network, or proteins in a protein-protein interaction network), and links representing interactions between the units (for example, friendship relationships or physical binding interactions between proteins). Network inference is the process of inferring the properties of those networks from data; typical network inference problems include the identification of groups of nodes with similar connection patterns, or the identification of unobserved interactions, that is, link prediction \cite{liben-nowell07,clauset08,guimera09,lu15,ghasemian20,guimera20}. Network inference and, in particular, link prediction are increasingly important in problems with applications ranging from the prediction of interactions between drugs \cite{guimera13,tarres19,menden19} to the prediction of human preferences and decisions \cite{guimera11,guimera12,godoylorite16,cobo18}.

Typically, network inference starts from observations of some of the links in the network, which are used to predict unobserved links or to infer other network properties. However, other sources of information such as system dynamics \cite{timme07,peixoto19} or node attributes \cite{tallberg04,yang13,hric16,newman16,white16,Peel17_,cobo18,stanley19,contisciani20} can also be used to aid in the inference process. Here we study how node attributes are introduced in the inference process, and what is the effect of using such metadata.

We present our work in terms of the problem of link prediction in recommender systems \cite{koren09,guimera12,godoylorite16}, in which the goal is to predict the association between users and items (for example, books or movies). However, our conclusions apply to network inference problems in general. We introduce a multipartite network model that encompasses and generalizes previous attempts to use node metadata in network inference problems (Fig.~\ref{fig:model_rep}). Within this framework, the problem of link prediction in general unipartite or bipartite networks is just a particular case. Unlike most previous approaches, our multipartite network model allows us to control the importance of the node metadata and thus to investigate when and how metadata helps in the inference.

We find that, contrary to what one may expect, node metadata do not affect the inference problem gradually. Rather, even when the weight of metadata increases smoothly, the inference process undergoes a transition from a situation in which metadata does not play any role, to a situation in which metadata completely dominates the inference process. When network data and metadata are partly correlated, metadata optimally contributes to the inference process at the transition between  data-dominated and  metadata-dominated regimes.

\section{Multipartipartite mixed-membership stochastic block models with labeled links}
\label{sect:model}
\begin{figure}[t]
	\centerline{
\includegraphics*[width=\columnwidth]{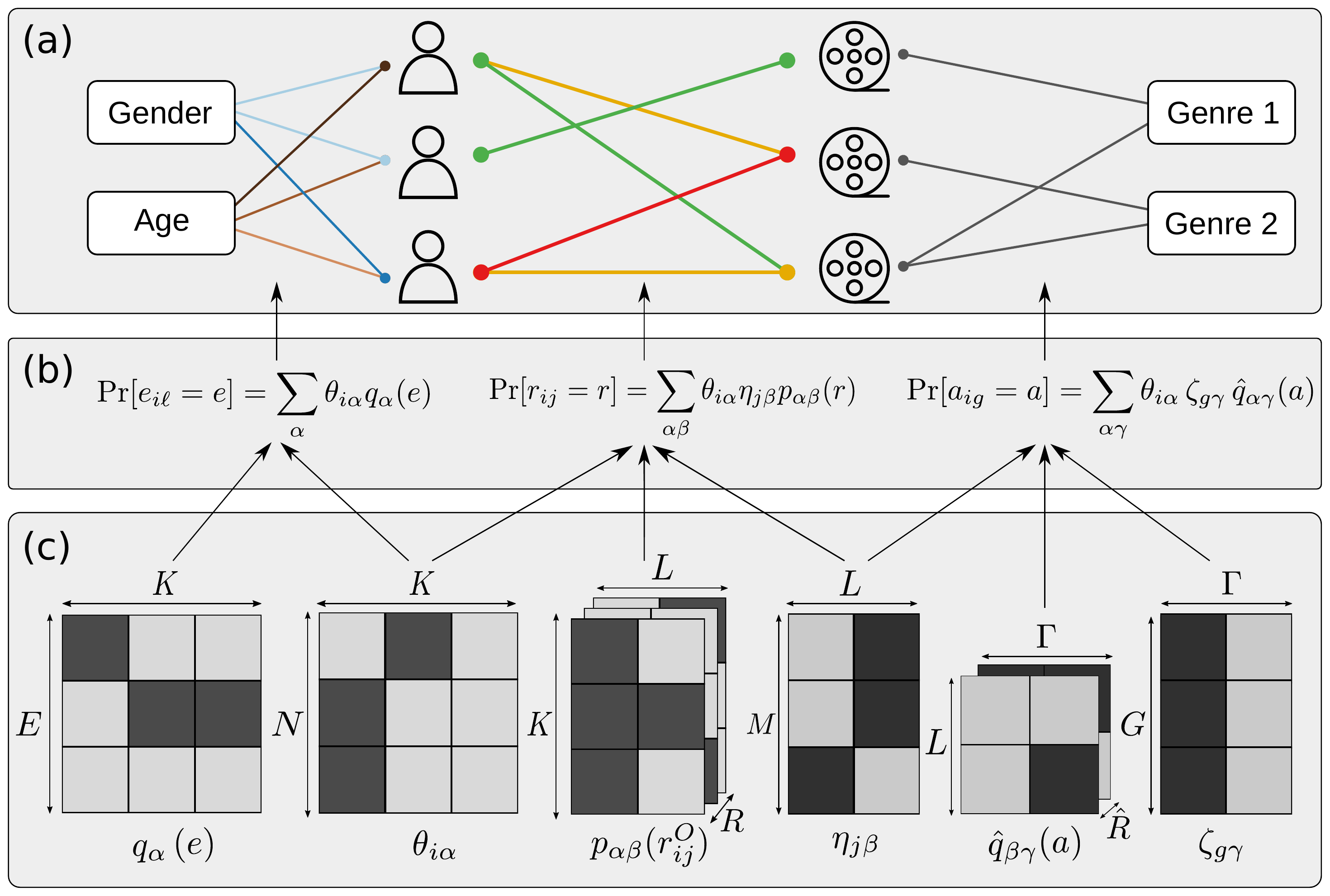}
}
	\caption{{\bf Multipartipartite mixed-membership stochastic block model with labeled links.}
	(a), We cast the recommendation problem (in which one aims to predict how users will rate certain items) into a network inference problem. Here, users rate movies with three possible ratings (green, orange or red). Additionally, we have excluding attributes for users (two excluding genders and three excluding age groups, represented by different shades of the same color) and non-excluding attributes for movies (two movie genres; the connection to these attributes is binary, yes/no, but in general it does not need to be). Similar to ratings, we represent these attributes as bipartite networks. Although we frame our description of the model in terms of recommendations or link prediction in a bipartite network, the problem of link prediction in regular unipartite networks is just a particular case in which user nodes and item nodes are the same.
	(b) Each bipartite network in the multipartite network is modeled using a mixed-membership stochastic block model (see text). The individual block models are coupled by the user and item membership vectors ($\thetas$ and $\etas$, respectively), shown in (c) along with all other model parameters and their dimensions (see text).
	}
	\label{fig:model_rep}
\end{figure}

We introduce a very general network model based on stochastic block models \cite{holland83,nowicki01,guimera09} that allows us to deal with (directed or undirected) unipartite and bipartite networks, whose links are binary or labeled, and with node attributes of different types that can be combined as needed (Fig.~\ref{fig:model_rep}). As we discuss below, this model extends and generalizes previous models.

In what follows we use the terminology of recommender systems \cite{koren09,guimera12,godoylorite16} although, as previously mentioned, the model is completely general and applicable to any type of relational data with node attributes. Our objective is to model a bipartite network with labeled links connecting $N$ users to $M$ items (for example, movies or books). Links $r_{ij}$ represent ratings of users $i$ to items $j$ and are labeled, that is, $r_{ij}$ can take values in a finite discrete set such as $\{{\rm like}, {\rm dislike}\}$, $\{{\rm green}, {\rm yellow}, {\rm red}\}$, or $\{0, 1, \dots, R\}$. To model these ratings, we assume that: (i) there are user and item groups, and users and items belong to mixtures of such groups; (ii) the probability that a user $i$ rates item $j$ with $r_{ij}$ depends only of the groups to which they belong.

These assumptions lead to a bipartite \cite {guimera11,guimera12,yen20} mixed-membership \cite{airoldi08} stochastic block model \cite{godoylorite16} in which the probability that user $i$ gives item $j$ a rating $r$ is
\begin{equation}
 \Pr[r_{ij} = r] = \sum_{\alpha\beta} \theta_{i\alpha} \eta_{j\beta} p_{\alpha\beta}(r) \;.
 \label{eq:link-model}
\end{equation}
Here, $\thetas_{i}$ is the normalized  membership vector of user $i$, and each element $\theta_{i\alpha}$ represents the probability that user $i$ belongs to group $\alpha$ (with $\sum_\alpha \theta_{i\alpha} =1$). Similarly, $\etas_{j}$ is the normalized membership vector of item $j$; $\eta_{j\beta}$ represents the probability that item $j$ belongs to group $\beta$. Finally, $p_{\alpha\beta}(r)$ is the probability that a user in group $\alpha$ and an item in group $\beta$ are connected with a rating $r$. The normalization condition here is $\sum_{r} p_{\alpha\beta}(r)=1$.

We note that the association between nodes (users and items)  and attributes can also be represented as a  bipartite network. Therefore we can model node-attribute  associations in a similar manner to ratings. Because we are interested in how node attributes can help in the inference of the model for ratings ($\thetas,\etas,{\bf p})$, we consider that membership vectors for users ($\thetas$) and items ($\etas$) in their respective attribute networks are the same as in the model for the ratings. 

We consider both {\em excluding} and {\em non-excluding attributes}. For excluding attributes, having one attribute excludes from having another; for example, a user's age group cannot be  30-39 years old and 40-49 years old simultaneously. We model each set of excluding attributes as a single attribute node (for example, an age node) that is connected to users or items through labeled links (each label representing a mutually excluding age group in the example). The probability that user $i$ has an excluding attribute $e$ (that is, the probability that the link $e_{i\ell}$ between user $i$ and attribute node $\ell$ is of type $e$) is
\begin{equation}
 \Pr[e_{i\ell} = e] = \sum_{\alpha} \theta_{i\alpha} q_{\alpha}(e) \;,
 \label{eq:link-model-nodes}
\end{equation}
where $q_{\alpha}(e)$ is the probability that a user of group $\alpha$ has an attribute of type  $e$, and $\sum_e q_\alpha(e)=1$. For items, the expression is identical except that we use item membership vectors $\eta$ instead of user membership vectors $\theta$. 

We also consider non-excluding attributes, such as item genre (for example, a movie could be both ``action'' and ``western''). We model each of these non-excluding attribute types as individual attribute nodes connected to user or item nodes by links that are typically binary (either do or do not have the attribute) but that could in general be also labeled. Then, the probability that item $i$ has attribute $g$ of type $a$ is also modeled using a mixed-membership, bipartite stochastic block model
\begin{equation}
 \Pr[a_{ig} = a] = \sum_{\alpha\gamma} \theta_{i\alpha} \, \zeta_{g\gamma} \,\hat{ q}_{\alpha\gamma}(a) 
 \label{eq:link-model-items}
\end{equation}
where $\zeta_{g\gamma}$ is the membership vector of attribute $g$ and $\hat{q}_{\alpha\gamma}(a)$ is the probability that a user in group $\alpha$ has an attribute of type $a$ for an attribute in attribute group $\gamma$. As before, the expression for item non-excluding attributes is identical, just replacing user membership vectors $\thetas$ by item membership vectors $\etas$.

\section{Model posterior and inference}
\label{sect:EM}

Our objective is to model the observed ratings $R^O$, and to predict the value of some unobserved ratings $R$. For this, and given Eq.~\eqref{eq:link-model}, we need to infer the parameters  $\thetas, \etas$ and $\ps$ from $R^O$; the posterior distribution over these parameters is given by
\begin{eqnarray}
 P( \boldsymbol{\theta},\boldsymbol{\eta},\boldsymbol{p}|R^O) & \propto &  P(R^O| \boldsymbol{\theta},\boldsymbol{\eta},\boldsymbol{p}) \, P( \boldsymbol{\theta},\boldsymbol{\eta},\boldsymbol{p}) \nonumber \\
 & \equiv & L^R(\boldsymbol{\theta},\boldsymbol{\eta},\boldsymbol{p}) \, P( \boldsymbol{\theta},\boldsymbol{\eta},\boldsymbol{p})\;,
 \label{eq:bayes}
\end{eqnarray}
where $L^R(\boldsymbol{\theta},\boldsymbol{\eta},\boldsymbol{p}) = P(R^O| \boldsymbol{\theta},\boldsymbol{\eta},\boldsymbol{p})$ is the likelihood of the model and $P( \boldsymbol{\theta},\boldsymbol{\eta},\boldsymbol{p})$ is the prior over model parameters. According to Eq.~\eqref{eq:link-model}, the likelihood is
\begin{equation}
L^R(\boldsymbol{\theta},\boldsymbol{\eta},\boldsymbol{p}) = \prod_{(i,j) \in R^O} \left[ \sum_{\alpha \beta} \theta_{i \alpha} \eta_{j \beta} p_{\alpha \beta}(r^O_{ij}) \right] \;. 
 \label{eq:likelihood}
\end{equation}

Similarly, if we decide to jointly model the ratings and the metadata encoded in the observed user and item attributes $A^O$, we also need to infer the values of the parameters $\zetas$, $\qs$ and $\hatqs)$ using the posterior
\begin{eqnarray}
 P( \boldsymbol{\theta},\boldsymbol{\eta},\boldsymbol{\zeta}, \boldsymbol{p}, \boldsymbol{q},  \hatqs|R^O, A^O) & \propto &  L^R(\boldsymbol{\theta},\boldsymbol{\eta},\boldsymbol{p}) \nonumber \times\\ & \times & \prod_k L^{A_k}(\boldsymbol{\theta},\boldsymbol{\eta},\boldsymbol{\zeta}, \boldsymbol{q},\hatqs) \times \nonumber\\
 &\times & P( \boldsymbol{\theta},\boldsymbol{\eta}, \boldsymbol{\zeta}, \boldsymbol{p}, \boldsymbol{q}, \hatqs)
 \label{eq:bayes2}
\end{eqnarray}
where $L^{A_k}(\boldsymbol{\theta},\boldsymbol{\eta},\boldsymbol{\zeta}, \boldsymbol{q},\hatqs)= P(A_k^O| \boldsymbol{\theta},\boldsymbol{\eta}, \boldsymbol{\zeta},\boldsymbol{q}, \hatqs)$ is the likelihood of the $k$-th attribute network (for example, the age attribute network for users, or the genre attribute network for items). For the $k$-th excluding attribute, this likelihood reads
\begin{equation}
 L^{A_k}(\thetas,\etas, \qs) = \prod_{(i, \ell_k) \in A_k^O} \left[ \sum_{\alpha} \theta_{i\alpha} q^k_{\alpha}((e_k^O)_{i\ell_k}) \right] \;,
 \label{eq:prior-people}
\end{equation}
where $\ell_k$ is the $k$-th non-excluding attribute and the product is over all nodes $i$ for which we observe attribute $\ell_k$.

For the $k$-th non excluding attribute we have
\begin{equation}
 L^{A_k}(\boldsymbol{\theta},\boldsymbol{\eta},\boldsymbol{\zeta}, \hatqs) = \prod_{(i, g) \in A_k^O} \left[ \sum_{\alpha\gamma} \theta_{i\alpha} \zeta^k_{g\gamma} \hat{q}^k_{\alpha\gamma}((a_k^O)_{ig}) \right] \;.
 \label{eq:prior-items}
\end{equation}
where the product is over all observed  associations  between nodes $i$ and attributes $g$ within the $k$-th  class of non-excluding attributes.

Ignoring normalizing constants, and in a spirit similar to Refs.~\cite{yang13,contisciani20}, we define a parametric log-posterior as
\begin{eqnarray}
 \pi( \boldsymbol{\theta},\boldsymbol{\eta},\boldsymbol{\zeta}, \boldsymbol{p}, \boldsymbol{q},\hatqs |R^O, A^O) & = &  \mathcal{L}^R(\boldsymbol{\theta},\boldsymbol{\eta},\boldsymbol{p}) + \nonumber\\
 & + & \sum_k \lambda_k \mathcal{L}^{A_k}(\boldsymbol{\theta},\boldsymbol{\eta},\boldsymbol{\zeta}, \boldsymbol{q}, \hatqs)\; ,
 \label{eq:posterior}
\end{eqnarray}
where $\mathcal{L}^R(\thetas, \etas,\ps)$ and $\mathcal{L}^{A_k}(\thetas, \etas, \zetas, \qs, \hatqs)$ are the log-likelihoods of ratings and attributes, respectively. For $\lambda_k=0$, we recover Eq.~\eqref{eq:bayes} with uniform priors on the parameters, thus completely ignoring all metadata. Conversely, for $\lambda_k=1$, we are jointly modeling the network of ratings and the network of attributes as in Eq.~\eqref{eq:bayes2}, with uniform priors on the parameters. By tuning the values of $\lambda_k$ we can interpolate between these situations, and extrapolate to situations with $\lambda_k > 1$ in which we would eventually only model the attribute network ($\lambda_k \gg 1$). The terms corresponding to the attribute models can indistinctly be interpreted as part of the likelihood of a joint model of ratings and attributes, similar to Refs.~\cite{yang13,hric16,stanley19,contisciani20}, or as a non-uniform prior over membership vectors as in Refs.~\cite{tallberg04,newman16,white16}. If interpreted as part of a joint model, then  $\lambda_k$ can be seen as some factors that are needed because attribute data are somehow less (or more) reliable than rating data, perhaps because we have reason to believe that attributes are more (or less) subject to noise, or because each rating corresponds, in fact, to a mean over several observations. Conversely, if interpreted as priors over the partitions,  $\lambda_k$ should be interpret as hyperparameters defining how certain we are a priori about the importance of node attributes.

Either way, this parametrized posterior allows us to investigate how the metadata encoded in the attribute networks enter the inference process for the ratings, and under which conditions it results in better and more predictive models for those ratings. To do this, we maximize the posterior for fixed values of $\lambda_k$ using an expectation-maximization algorithm \cite{godoylorite16,newman16,stanley19,contisciani20} (see Appendix \ref{app:em}), which gives the most plausible parameter values. Because the posterior landscape is in general rugged, we perform several runs of the EM algorithm and compute the average probability for each unobserved rating to make predictions (see \cite{godoylorite16} and Appendix \ref{app:em}).

\section{Relationship to previous work}

The literature on using metadata for link prediction and recommender systems is vast, and includes all sort of approaches ranging from simple heuristics to sophisticated machine learning methods. However, our interest here is more closely related to probabilistic approaches to network inference, even when those approaches are not applied directly to link prediction \cite{tallberg04,yang13,hric16,newman16,white16,Peel17_,cobo18}---as shown in Refs.~\cite{stanley19,contisciani20}, once model parameters are inferred for, for example, community detection, they can easily be used to predict links as well. Our focus on approaches based on probabilistic generative models is motivated by three characteristics of such approaches: (i) all assumptions in them are explicit; (ii) principled (as opposed to heuristic) and sometimes even exact inference approaches are possible; and (iii) their results are more readily interpretable. These three characteristics make probabilistic approaches especially appropriate for our ultimate goal of understanding how node attributes enter and help in the inference process.

From this perspective, the multipartite mixed-membership stochastic block model is useful because it extends and generalizes previous models. By introducing excluding and non-excluding attributes, the model can accommodate simultaneously attributes like those considered in Refs.~\cite{newman16,contisciani20} (excluding) and in Refs.~\cite{yang13,hric16} (non-excluding). It can also combine an arbitrary number of attributes of different types, unlike approaches that can only deal with single attributes \cite{newman16,contisciani20} or, more often, with a single type of attribute; and it deals naturally with missing attribute data, unlike approaches that require all node attributes to be known \cite{tallberg04,white16}. Since attributes are modeled with a stochastic block model, our approach also automatically clusters attributes that have similar effects on the data (for example, age groups that show similar behavior) as in Ref. \cite{hric16}. Unlike most previous approaches for attributed networks, nodes and attributes in our model belong to mixtures of groups, which makes the model more expressive \cite{godoylorite16}, links between nodes and to attributes can be labeled, and the influence of the attributes can be tuned on and off (as in Ref.~\cite{contisciani20}). As stated above, this last feature is precisely the main focus of our work.

\section{Synthetic data}

We first use synthetic data to validate the expectation-maximization inference approach and to investigate the role of introducing node attributes. We generate synthetic data with a model similar to the model Fig.~\ref{fig:model_rep}. Our synthetic rating networks consist of 200 users and 200 items, partitioned into $K=2$ groups of users and  $L=4$ groups of items. Users have an excluding attribute labeled ``male'' or ``female'', and items have an excluding attribute labeled from 0 to 3, which may represent four different genres.

In the simplest case, in which ratings and attributes are completely correlated, all female users have membership vectors $\thetas_{\rm f} = (0.8, 0.2)$; conversely, all male users have $\thetas_{\rm m} = (0.2, 0.8)$. Similarly, an item with attribute $a$ has a membership of 0.8 to group $a$ and 0.067 to all other groups. 
To simulate partial correlation $c$ or even no correlation ($c=0$) between membership vectors and attributes, with probability $1-c$ we reassign each node attribute to a value selected uniformly at random among all possibilities (2 for users and 4 for items).

\begin{figure}[t!]
	\centerline{
\includegraphics*[width=.5\textwidth]{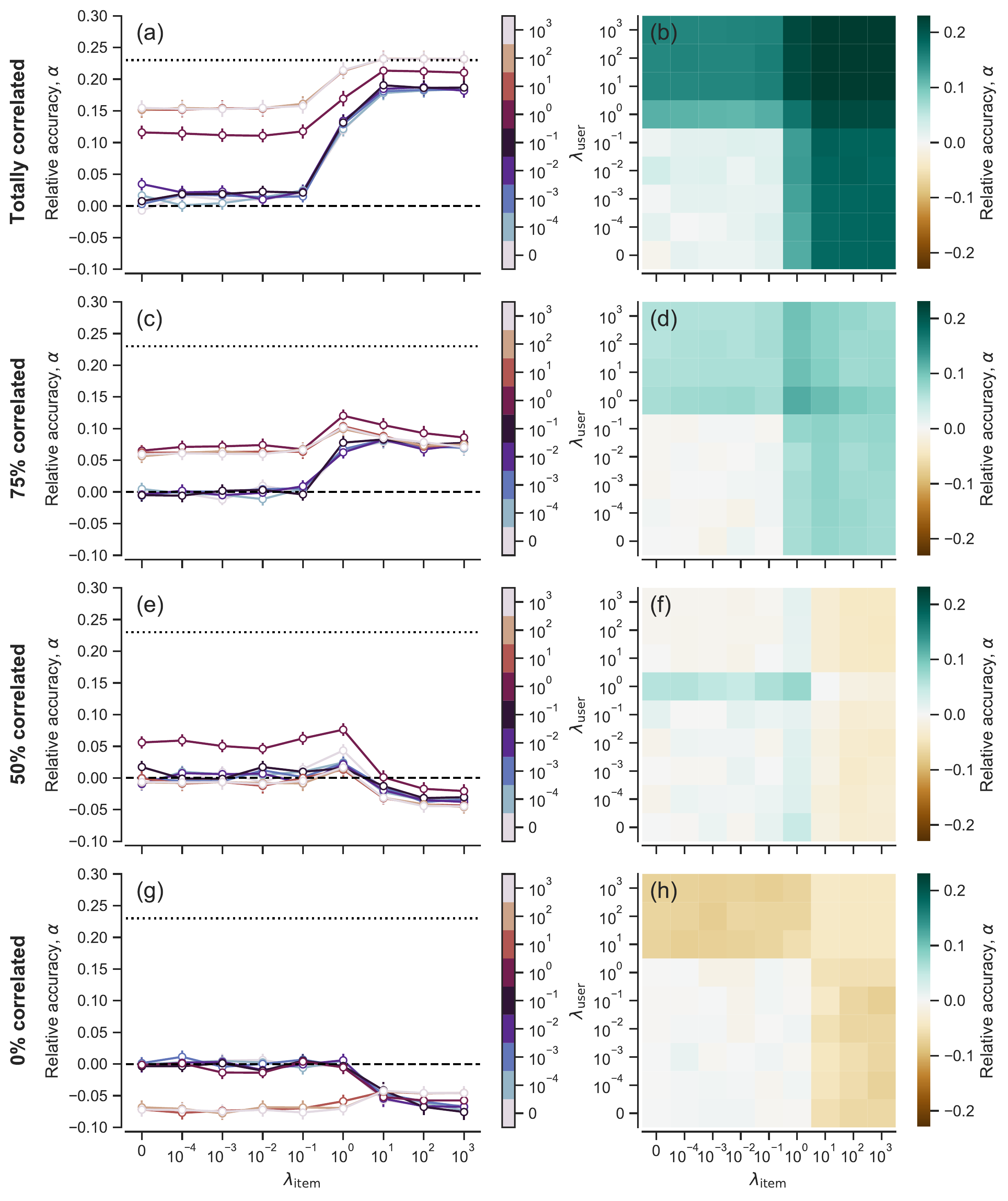}
}
	\caption{{\bf Predictive performance and effect of metadata on synthetic ratings.}
	We create synthetic ratings from 200 users on 200 items, with different levels of correlation $c$ between ratings and node attributes (see text). We then use 5-fold cross-validation to calculate the performance of the expectation-maximization equations at predicting unobserved ratings. In particular, we take as a reference the predictive accuracy $\ac_0$ of the algorithm when all attributes are ignored ($\lambda_{\rm user}=\lambda_{\rm item}=0$), and measure relative accuracy $\alpha$ for a given pair $(\lambda_{\rm user}, \lambda_{\rm item})$ as the log-ratio  
    $\alpha(\lambda_{\rm user}, \lambda_{\rm item}) = \log \left[\ac(\lambda_{\rm user}, \lambda_{\rm item}) / \ac_0\right]$. The value $\alpha(\lambda_{\rm user}, \lambda_{\rm item}) = 0$ (dashed line) thus indicates no change with respect to the reference $\ac_0$, and $\alpha(\lambda_{\rm user}, \lambda_{\rm item}) > 0$ (respectively, $\alpha(\lambda_{\rm user}, \lambda_{\rm item}) < 0$) indicates predictions that are more (less) accurate than those obtained by ignoring node attributes. The maximum possible relative performance (dotted line) is obtained when each rating is assigned the exact probability that was used to generate it.
	For each value of the correlation ((a)-(b), full correlation, $c=1$; (c)-(d), $c=0.75$; (e)-(f), $c=0.50$; (g)-(h), no correlation, $c=0$) we show the variation of $\alpha(\lambda_{\rm user}, \lambda_{\rm item})$ with $\lambda_{\rm item}$ for different values of $\lambda_{\rm user}$ (left), and the whole dependence of $\alpha(\lambda_{\rm user}, \lambda_{\rm item})$ on both $\lambda_{\rm user}$ and $\lambda_{\rm item}$ (right).
	}
	\label{fig:synthetic}
\end{figure}

For the experiments reported in Fig.~\ref{fig:synthetic}, we consider all attribute links, but only a number $|R^O|=400$ of observed ratings (that is, $1\%$ of all generated ratings). Although the synthetic data are created with item genre as an excluding attribute, we carry out the inference process assuming that genre is a non-excluding attribute, which is what one would likely assume in real settings where the generating model is unknown.

We infer the values of the model parameters using the expectation-maximization equations, and use the inferred parameters to predict unobserved ratings in the bipartite ratings network. We do this for different levels of correlation $c$ between the ratings and the attribute networks (Fig.~\ref{fig:synthetic}), from a situation $c=1$ in which the attributes are perfectly correlated with user and item membership vectors (all male users belong to one group and have identical parameters, and all females belong to another group with different parameters; items with each genre belong to the exact same mixture of groups) to a situation $c=0$ in which user and item memberships and attributes are completely uncorrelated (Fig.~\ref{fig:synthetic}).

Since we focus on sparse observations in which the number of observed ratings is low (only $1\%$ of all ratings), model parameters cannot be inferred accurately from the ratings alone. Therefore, when we only consider the observed ratings $R^O$ and ignore all attributes $A^O$ by setting $\lambda_{\rm user}=\lambda_{\rm item}=0$ in Eq.~\eqref{eq:posterior} ($\lambda_{\rm user}$ and $\lambda_{\rm item}$ correspond to the user and item attribute networks, respectively), the prediction of unobserved links is suboptimal, that is, the inferred probabilities of unobserved links differ significantly from the actual probabilities used to build the network.

When there is perfect correlation between node attributes and group memberships, considering the attributes $A^O$ by setting $\lambda_{\rm user}>0$ and $\lambda_{\rm item}>0$ should in principle help in the inference process. In fact, since attributes are perfectly correlated to group memberships, in the limit $\lambda_{\rm user}\rightarrow\infty$ and $\lambda_{\rm item}\rightarrow\infty$ nodes will be forced into the correct groups and predictions should be near optimal. This is what we observe in our numerical experiments (Fig.~\ref{fig:synthetic}a). Interestingly, as we increase the weight of the  attributes in the log-posterior  from $\lambda_{\rm user}=\lambda_{\rm item}=0$, the effect on prediction accuracy is not smooth. Rather, below certain threshold values of $\lambda_{\rm user}$ and $\lambda_{\rm item}$, using the attributes does not have any significant effect on prediction accuracy. Then, at those threshold values, a transition occurs and prediction accuracy increases abruptly until it reaches its theoretical maximum, as expected.

When attributes and ratings are completely uncorrelated (Fig.~\ref{fig:synthetic}d), the role of attributes is reversed. Predictions are equally suboptimal at $\lambda_{\rm user}=\lambda_{\rm item}=0$, but then, as $\lambda_{\rm user}$ and $\lambda_{\rm item}$ cross certain threshold values, predictions suddenly worsen as user and item nodes are forced into groups that are uncorrelated with their real membership vectors and, thus, with the observed ratings.

\begin{figure}[]
	\centerline{
    \includegraphics*[width=\columnwidth]{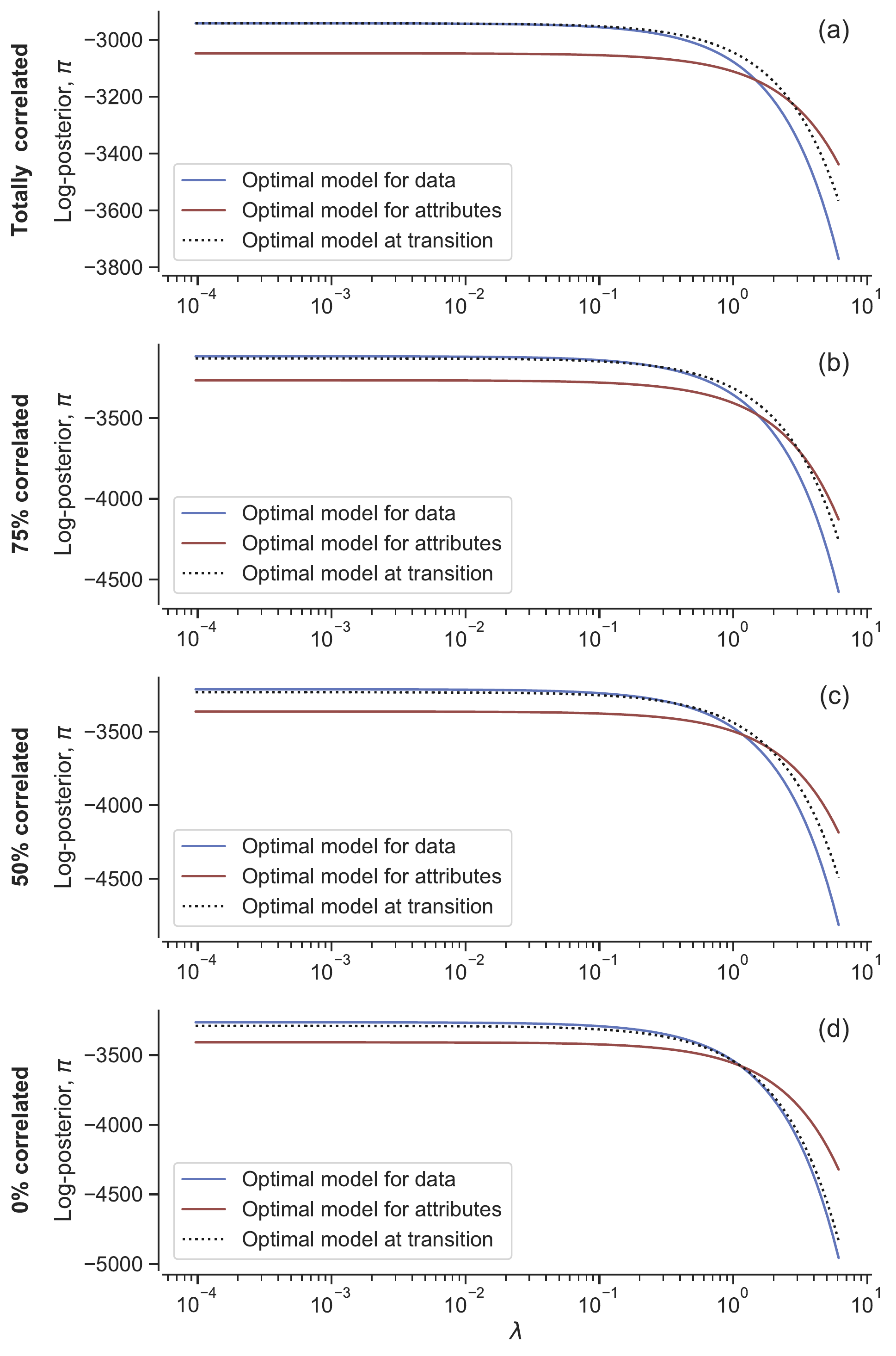}}
	\caption{{\bf Transition between data-dominated and metadata-dominated inference regimes.}
	For the synthetic data in Fig.~\ref{fig:synthetic}, we plot the log-posterior $\pi( \thetas,\etas,\zetas, \ps, \qs, \hatqs |R^O, A^O)$ as a function of the hyperparameter $\lambda=\lambda_{\rm item}=\lambda_{\rm user}$ for three models: the model that maximizes the data likelihood $L^R$, the model that maximizes the metadata likelihood $L^A$, and the model that maximizes the posterior when two previous cases cross (that is, have equal posteriors). The position of the crossing coincides with the transitions and the maxima observed in Fig.~\ref{fig:synthetic}.
	}
	\label{fig:crossing}
\end{figure}

Unlike the extreme cases of total correlation or zero correlation, when attributes are partly correlated with the true group memberships of the nodes, the change in performance is not monotonic as we increase the importance of the attributes. As before, when $\lambda_{\rm user}$ and $\lambda_{\rm item}$ are small enough, we observe no difference with the situation in which the attributes are ignored entirely. In the other extreme, when $\lambda_{\rm user}\rightarrow\infty$ and $\lambda_{\rm item}\rightarrow\infty$ user and item nodes are forced into groups that match partly, but not perfectly, the true group memberships of the nodes, so the performance may increase or decrease with respect to the situation with no attributes, depending on whether the correlation is high (Fig.~\ref{fig:synthetic}b) or low (Fig.~\ref{fig:synthetic}c). However, we find that the most predictive models in this case are those at intermediate values of $\lambda_{\rm user}$ and $\lambda_{\rm item}$, precisely at the transition region where both the observed ratings and the observed attributes play a role in determining the most plausible group memberships. In this case, the inferred node memberships do not coincide with either those that maximize $L^R$ of those that maximize $L^{A_k}$.

To understand the transition from the rating-dominated to the attribute-dominated regime, we study the posterior of the two extreme models corresponding to the maximum a posterior estimates obtained by expectation-maximization for $\lambda_{\rm user}=\lambda_{\rm item}=0$ and for $\lambda_{\rm user}=\lambda_{\rm item}\rightarrow\infty$ (Fig.~\ref{fig:crossing}). These are the most plausible models when only data (ratings) and only metadata (attributes) are taken into consideration, respectively. Regardless of the correlation between ratings and attributes, we find that the transition in predictability in Fig.~\ref{fig:synthetic} coincides with the region where the data-dominated and metadata-dominated posteriors cross. By considering Eq.~\eqref{eq:posterior} we see that this must be the case. Indeed, for each attribute network we find three regimes---one dominated by the $L^R$ term, one dominated by the $L^A$ term, and one in which both terms are comparable. Unless there is perfect or almost perfect correlation between attributes and node memberships, any improvement in predictive power must come from considering both the observed ratings and the observed attributes, and therefore in the transition region.

\section{Real data}

\begin{figure}[t!]
	\centerline{
    \includegraphics*[width=\columnwidth]{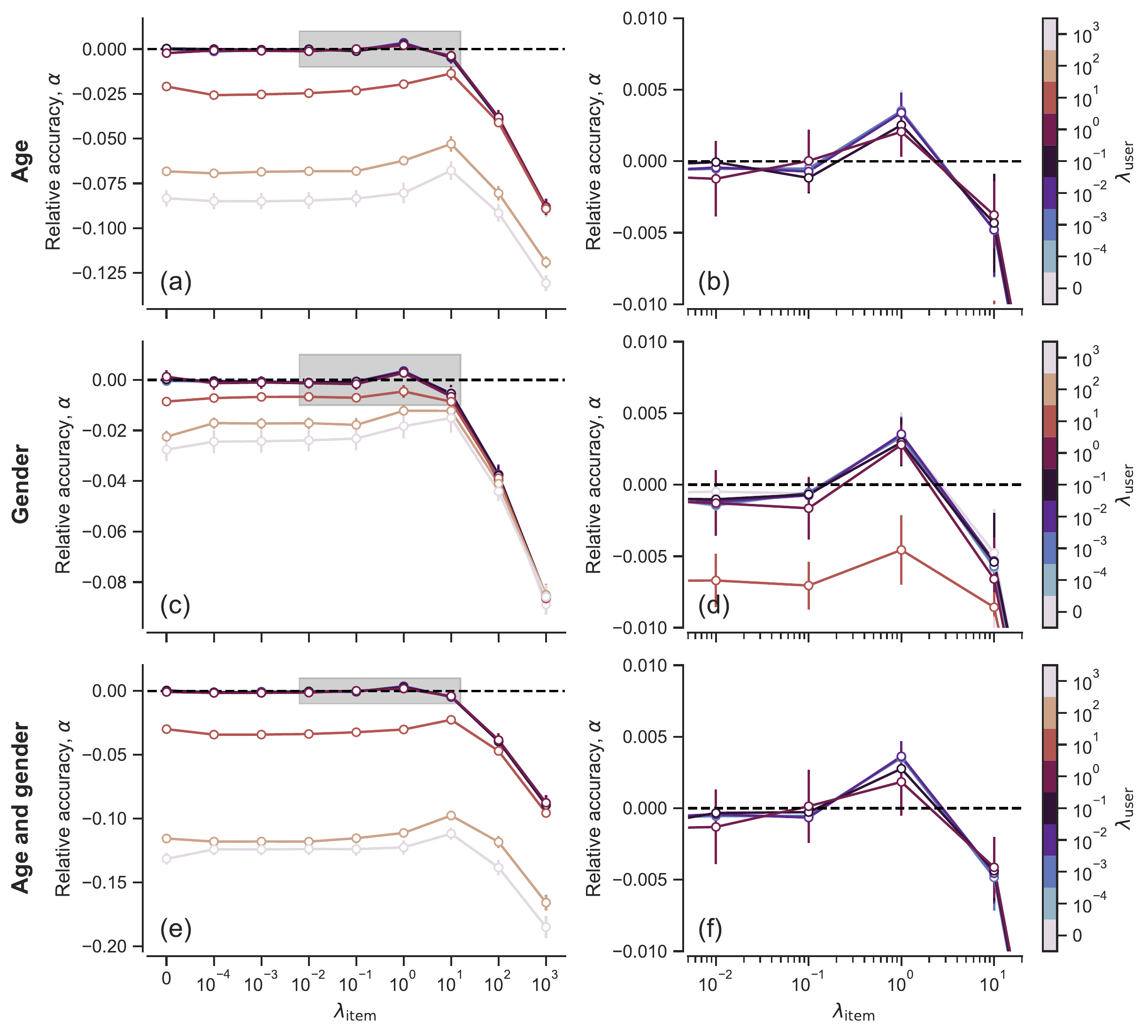}
    }
	\caption{{\bf Predictive performance and effect of metadata on the MovieLens data set.}
	As in Fig.~\ref{fig:synthetic}, we take as a reference the predictive accuracy $\ac_0$ of the algorithm when all attributes are ignored ($\lambda_{\rm user}=\lambda_{\rm item}=0$), and measure relative accuracy $\alpha$ for a given pair $(\lambda_{\rm user}, \lambda_{\rm item})$ as the log-ratio $\alpha(\lambda_{\rm user}, \lambda_{\rm item}) = \log \left[\ac(\lambda_{\rm user}, \lambda_{\rm item}) / \ac_0\right]$.
	We consider three different attributes for user nodes: (a)-(b), age; (c)-(d), gender; (e)-(f), age and gender combined as a single attribute. We plot the whole range of $\lambda_{\rm user}$ (left), and zoom into the intermediate (shaded) region of $\lambda_{\rm user}$ in which predictions are significantly more accurate than the reference (right).
	}
	\label{fig:movielens}
\end{figure}

Finally, we analyze two empirical data sets and study whether we observe the same behaviors as in the synthetic data. First, we consider the 100K MovieLens data set \cite{movielens}, which contains 100,000 ratings of movies by users. Age and gender attributes are available for users, which we model as excluding attributes (Fig.~\ref{fig:movielens}). Movies have genre attributes, which we model as non-excluding attributes. The relative weights of user and movie attributes are given by the parameters $\lambda_{\rm users}$ and $\lambda_{\rm items}$.

Just as in the synthetic networks with small but finite correlation, we observe an intermediate value of $\lambda_{\rm user}$ and $\lambda_{\rm item}$ that provides more accurate rating predictions than either considering the observed ratings alone or considering the node attributes alone. This behavior is similar when we consider age only, gender only, or age and gender simultaneously. As in synthetic networks, the optimal combination of rating data and node metadata occurs for values of $\lambda$ such that the ratings network and the attributes networks have comparable contributions to the log-posterior. 

\begin{figure}[t!]
	\centerline{\includegraphics*[width=.5\textwidth]{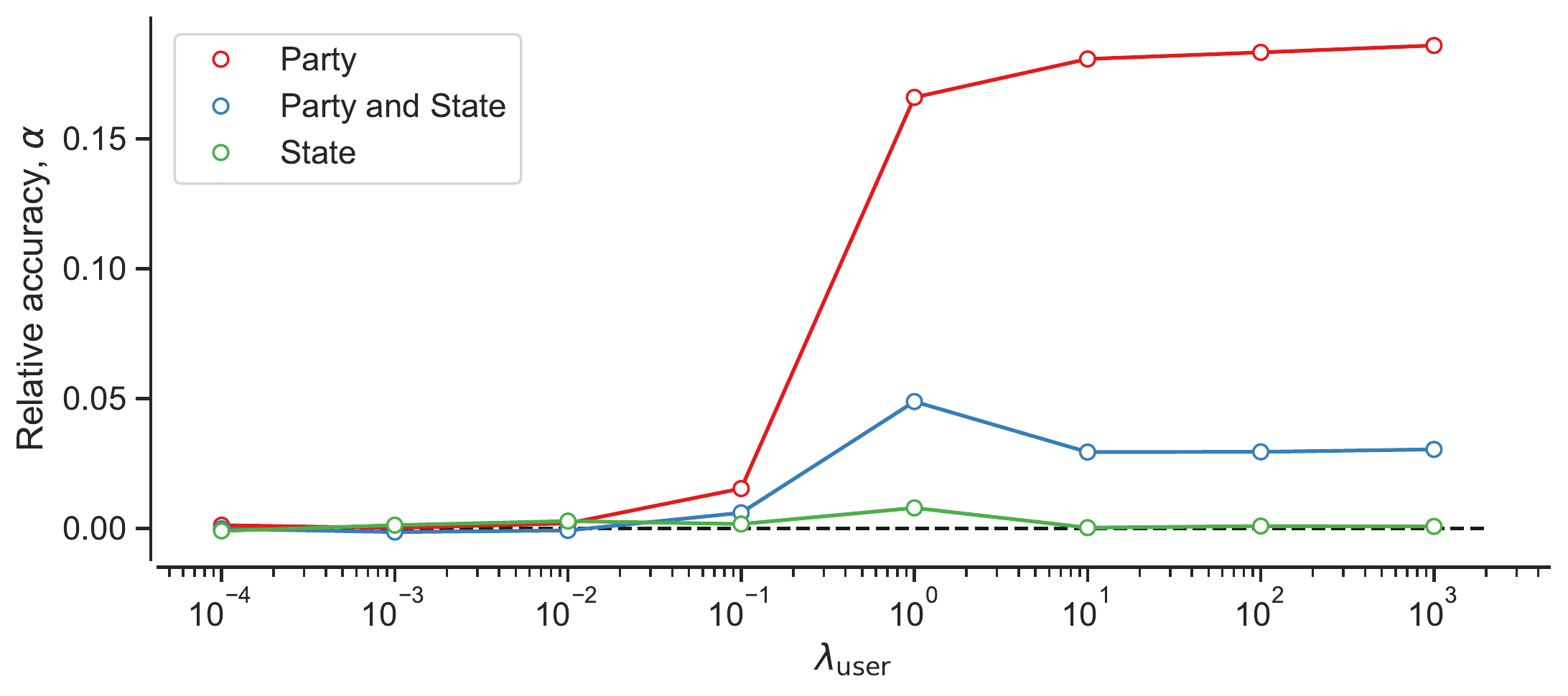}
}
	\caption{{\bf Predictive performance and effect of metadata on the U.S. Congress data set.}
	As in Fig.~\ref{fig:synthetic}, we take as a reference the predictive accuracy $\ac_0$ of the algorithm when all attributes are ignored ($\lambda_{\rm user}=0$), and measure relative accuracy $\alpha$ for a given $\lambda_{\rm user}$ as the log-ratio $\alpha(\lambda_{\rm user}) = \log \left[\ac(\lambda_{\rm user}) / \ac_0\right]$.
	We consider three different attributes for user nodes: Party, State, and party and State simultaneously.
	}
	\label{fig:congress}
\end{figure}

Second, we consider a data set on the votes of 441 members of the U.S. House of Representatives in the 108th U.S. Congress \cite{waugh09} (Fig.~\ref{fig:congress}). Between Jannuary 2003 and Jannuary 2005, these representatives voted on 1,217 bills, casting one of 9 different types of vote, which, following previous analyses, we simplify to Yes, No, and Other \cite{waugh09}.
In this data set, ``users'' are the representatives and ``items'' are the bills. The ratings represent the votes of the representatives on the bills. For representatives, we have attribute data indicating their party and state, which we model as excluding attributes. Although all votes of all members are recorded in the data set (in total, 536,698 votes), for the purpose of our analysis we infer the parameters of the multipartite mixed-membership stochastic block model using 1\% of the data, and predict the remaining 99\% (and repeat this using each 1\% of the data as training set).

Again, the effects of introducing the attributes in the inference process are very similar to those we encounter in synthetic data (Fig.~\ref{fig:congress}). When using only the state of the representatives, we observe a behavior that is compatible with small but finite correlation between attribute and voting patterns, since the optimal predictive performance is observed at intermediate values of $\lambda_{\rm user}$. Rather, when we consider party affiliation we observe a behavior that is compatible with almost perfect correlation between attribute and voting behavior. Indeed, in this case the predictive performance of the model increases monotonically with $\lambda_{\rm user}$, with an abrupt transition at $\lambda_{\rm user} \approx 1$, just as for perfectly correlated attributes in synthetic data. When state and party are combined into a single excluding attribute (for example, ``Democrat from Texas'' is a group), we observe a behavior compatible with strong (but imperfect) correlation between attributes and voting behavior. In this case, predictive accuracy does not improve monotonically with $\lambda_{\rm user}$ because, for very large values, representatives are forced into small groups that are more prone to fluctuations, that is, the model overfits the data thus worsening the predictive power with respect to considering large groups associated to party affiliation alone.

\section{Conclusion}

There is ample evidence that using node metadata can help to solve network inference problems. As we have discussed, several approaches have been proposed in recent years to introduce node attributes into probabilistic network models, and to use them to make better inferences about, for example, the group structure of networks or the existence of unobserved interactions. In these approaches, node attributes are introduced either as part of a whole-system model (including both the links between nodes and node attributes), or as priors over the parameters of the model for the links (for example, as priors for the node group memberships that, in turn, determine the probability of existence of links). However, beyond the improvement in performance that they may entail in a given task such as group detection or link prediction, we know little about the effect that node attributes have in the inference process. Here, our goal has been to clarify this issue.

Regardless of whether attributes are introduced as part of a whole model or as a prior for model parameters, they appear in probabilistic models as additional terms in the likelihood or the posterior. As we have shown, our results depend on this simple observation alone---only when all terms in these likelihoods or posteriors are comparable in magnitude, or when attributes are perfectly correlated with ratings, can we expect attributes to improve the inference process. In this sense, our findings here may be expected to be universal.

From a practical point of view, our work helps to understand when certain approaches will not work. For example, our results suggest that modeling data and metadata jointly will only improve link predictions (or other network inference problems) if two conditions are fulfilled simultaneously: (i) the metadata are correlated to the data; (ii) as we have mentioned, the balance between amount of data and metadata is such that their likelihoods ($L^R$ and $L^A$ above) are of the same order. If the first condition is not fulfilled, using metadata will in general worsen predictions, rather than improving them; if the second condition is not fulfilled, one may, in practice, inadvertently ignore either the data or the metadata and thus make, again, suboptimal predictions.

Some works have intuitively addressed this problem by introducing tuning parameters akin to our $\lambda_k$ \cite{yang13,contisciani20}. However, the impact of those parameters has not been studied in detail and, instead, their values are typically chosen among a very limited set by means of cross-validation. Our work clarifies how the value of those parameters should be chosen, and why.

From a broader perspective, our work opens the door to understanding the role of different terms in probabilistic network models, as well as the transitions that occur between the regimes in which one term or another dominates. This sets the stage for more systematic approaches to building better probabilistic models of network systems.

\begin{acknowledgements}
The authors acknowledge support by the Spanish Ministerio de Econom\'{\i}a y Competitividad (Grants FIS2016-78904-C3-P-1 and PID2019-106811GB-C31) and by the Government of Catalonia (Grant 2017SGR-896).
\end{acknowledgements}

\appendix
\section{Expectation-maximization equations}
\label{app:em}

We aim to maximize the parametric log-posterior in Eq.~\eqref{eq:posterior} as a function of the model parameters $\thetas, \etas, \ps, \zetas ,\qs$ and $ \hatqs$. Because logarithms of sums are hard to deal with, we use a variational trick that first introduces an auxiliary distribution $p(x)$ with  $ \sum_x p(x) =1$ into a sum of terms as $\sum_x x  = \sum_x p(x) \left (x/ p(x) \right)$. Then because $\sum_x p(x) \left (x/ p(x) \right)= \langle x/p(x)\rangle$ we can use Jensens' inequality $\log \langle y\rangle \geq \langle \log y \rangle $ to write 
$\log \left[ \sum_x p(x) \left( x/ p(x) \right) \right] \geq \sum_x p(x) \log \left[ x/p(x) \right]$. 

Because both rating and attribute terms in Eq.~\eqref{eq:posterior} contain logarithms of sums, we introduce an auxiliary distribution for each of the terms as follows. For the ratings, we have
\begin{eqnarray}
 \mathcal{L}^R & = &  \sum_{(i,j) \in R^O} \log \sum_{\alpha \beta} \theta_{i \alpha} \eta_{j \beta} p_{\alpha \beta}(r^O_{ij}) r \nonumber \\ 
 & = & \sum_{(i,j) \in R^O} \log  \sum_{\alpha \beta} \omega_{ij}(\alpha,\beta)\frac{\theta_{i \alpha} \eta_{j \beta} p_{\alpha \beta}(r^O_{ij})}{\omega_{ij}(\alpha,\beta)}   \nonumber\\
 & \geq & \sum_{(i,j) \in R^O} \sum_{\alpha \beta}  \omega_{ij}(\alpha,\beta)\log \frac{\theta_{i \alpha} \eta_{j \beta} p_{\alpha \beta}(r^O_{ij})}{\omega_{ij}(\alpha,\beta)}
  \label{eq:posterior1}
  \end{eqnarray}
  where $\omega_{ij}(\alpha,\beta)$ is the auxiliary distribution.
  
 For the term corresponding to excluding node attributes we have
 \begin{eqnarray}
 \mathcal{L}^{A_k}  & = &   \sum_{(i, \ell_k) \in A_k^O} \log  \sum_{\alpha} \theta_{i\alpha} q_{\alpha}^k(i\ell_k) \nonumber\\ 
 & = &  \sum_{(i, \ell_k) \in A_k^O} \log  \sum_{\alpha}\sigma_{i\ell_k}^k(\alpha) \frac{\theta_{i\alpha} q_{\alpha}^k(i\ell_k)}{\sigma_{i\ell_k}(\alpha)} \nonumber\\
 & \geq & \sum_{(i, \ell_k) \in A_k^O} \sum_{\alpha}\sigma_{i\ell_k}^k(\alpha) \log \frac{\theta_{i\alpha} q_{\alpha}^k(i\ell_k)}{\sigma_{i\ell_k}^k(\alpha)} \label{eq:posterior2}
\end{eqnarray}
where $ \sigma^k_{i\ell_k}(\alpha)$ is the auxiliary distribution, and to simplify  the notation we have defined $q^k_{\alpha}(i\ell_k) \equiv q^k_{\alpha}(\left( e_k^O \right)_{i\ell_k})$.

Finally, for the term corresponding to non-excluding node attributes we have
\begin{eqnarray}
 \mathcal{L}^{A_k} & = & \sum_{(i, g) \in A_k^O}  \log \sum_{\alpha\gamma} \theta_{i\alpha} \zeta_{g\gamma}^k \hat{q}_{\alpha\gamma}(ig) \nonumber \\ 
 & = & \sum_{(i, g) \in A_k^O}  \log \sum_{\alpha\gamma} \hat{\sigma}_{ig}^k(\alpha,\gamma) \frac{\theta_{i\alpha} \zeta_{g\gamma}^k \hat{q}_{\alpha\gamma}(ig)}{\hat{\sigma}_{ig}^k(\alpha,\gamma)} \nonumber\\
 & \geq & \sum_{(i, g) \in A_k^O} \sum_{\alpha\gamma} \hat{\sigma}_{ig}^k(\alpha,\gamma) \log  \frac{\theta_{i\alpha} \zeta_{g\gamma}^k \hat{q}_{\alpha\gamma}(ig)}{\hat{\sigma}_{ig}^k(\alpha,\gamma)}
 \label{eq:posterior3}
\end{eqnarray}
where $\hat{\sigma}_{ig}^k(\alpha,\gamma)$ is the auxiliary distribution, and to simplify  the notation we have defined
 $\hat{q}^k_{\alpha}(ig)\equiv \hat{q}_{\alpha\gamma}^k(\left (a_k^O\right)_{ig})$.

Note that, in Eqs.~\eqref{eq:posterior1}-\eqref{eq:posterior3} above, the equality is satisfied when maximizing with respect to the auxiliary distributions. By solving these optimization problems we obtain
\begin{eqnarray}
 \omega_{ij}(\alpha,\beta) & = & \frac{ \theta_{i \alpha} \eta_{j \beta} p_{\alpha \beta}(r^O_{ij}) }{\sum_{\alpha' \beta'}\theta_{i \alpha'} \eta_{j \beta'} p_{\alpha' \beta'}(r^O_{ij})} \;, \\
 \sigma_{i\ell_k}^k(\alpha) & = & \frac{\theta_{i\alpha} q_{\alpha}^k(i\ell_k)}{\sum_{\alpha' }\theta_{i\alpha'} q^k_{\alpha'}(i\ell_k)}\;, \\
  \hat{\sigma}_{ig}^k(\alpha,\gamma) & = & \frac{\theta_{i\alpha} \zeta_{g\gamma}^k \hat{q}_{\alpha\gamma}(ig)}{\sum_{\alpha' \gamma'}\theta_{i\alpha'} \zeta_{g\gamma'} \hat{q}_{\alpha'\gamma'}(ig)} \,.
\end{eqnarray}
Therefore, the auxiliary distributions have the following interpretations: $\omega_{ij}(\alpha,\beta)$ is the contribution of user group $\alpha$ and item group $\beta$ to the probability that user $i$ gives item $j$ a rating $r^O_{ij}$; $\sigma^k_{i\ell_k}(\alpha)$ is the contribution of user group (or item group) $\alpha$ to the probability that user (item) $i$ has attribute type  $(e_k^O)_{i\ell_k}$ in the $k$-th excluding attribute; and, finally, $\hat{\sigma}^k_{ig}(\alpha,\gamma)$ is the contribution of groups $\alpha$ and $\gamma$ to the probability that, for the $k$-th non-excluding attribute, the association between node $i$ and attribute $g$ is of type  $(a_k^O)_{ig}$.

Using Lagrange multipliers for the normalization constraints, and equating to zero the derivatives of the log-posterior with respect to the model parameters yields
\begin{widetext}
\begin{equation}
 \theta_{i\alpha} = \frac{\sum_{j \in \partial i}\sum_\beta  \omega_{ij}(\alpha,\beta) + \sum_k \lambda_k  \sigma^k_{i\ell_k}(\alpha) + \sum_l \lambda_l \sum_{g \in {\partial_i}^k}\sum_\gamma  \hat{\sigma}_{ig}^l(\alpha,\gamma)  }{d_i+\sum_k\lambda_k \delta_i^k+\sum_l \lambda_l \Delta_i^l}
\end{equation}
where $\partial_i^k$ is the set of $k$-th attributes associated with  user $i$, $d_i$ is the degree of user $i$ in the network of ratings, and   $\Delta_i^l =|{\partial_i}^l|$. Note that the term $ \sigma^k_{i\ell_k}(\alpha)$ is equal to zero if user $i$ does not have attribute $\ell_k$, so that $\delta_i^k=1$ if user $i$ has exclusive attribute $\ell_k$ and zero otherwise.

\begin{equation}
 \eta_{j\beta} = \frac{\sum_{i \in \partial j}\sum_\alpha  \omega_{ij}(\alpha,\beta) + \sum_k \lambda_k  \sigma^k_{j\ell_k}(\beta) + \sum_l \lambda_l \sum_{i \in \partial_j^k}\sum_\gamma \hat{\sigma}_{ij}^l(\beta,\gamma)  }{d_j+\sum_k \lambda_k \delta_j^k +\sum_l \lambda_l \Delta_j^l}
\end{equation}
\end{widetext}
where $\partial_j^k$ is the set of $k$-th attributes associated with  item $j$, $d_j$ is the degree of item $j$ in the network of ratings, and   $\Delta_j^l =|{\partial_j}^l|$. As before, the term $ \sigma^k_{j\ell_k}(\beta)$ is equal to zero if item $j$ does not have attribute $\ell_k$, so that $\delta_j^k=1$ ifitem $j$ has exclusive attribute $\ell_k$ and zero otherwise.

\begin{equation}
 \zeta_{g\gamma}^k =  \frac{ \sum_{i \in \partial_g^k}\sum_\alpha  \hat{\sigma}_{ig}^k(\alpha,\gamma)  }{  \Delta_g^k}
\end{equation}
where where $\partial_g^k$ is the set of nodes associated with attribute $g$, and   $\Delta_g^k =|{\partial_g}^k|$.
Additionally, we have
\begin{equation}
    p_{\alpha\beta}(r)=\frac{\sum_{(i,j) \in R^O | r_{ij}^0}=r \omega_{ij}(\alpha,\beta)}{\sum_{(i,j) \in R^O}\omega_{ij}(\alpha,\beta)}
\end{equation}

\begin{equation}
    q_{\alpha}^k(e)=\frac{\sum_{(i,\ell_k) \in A_k^O| (e^O_k)_{i\ell_k}=e} \sigma_{i\ell_k}^k(\alpha)}{\sum_{(i,\ell_k)}  \sigma_{i\ell_k}^k(\alpha)}
\end{equation}
\begin{equation}
    \hat{q}_{\alpha\gamma}^k(a)=\frac{\sum_{(i,g) \in A_k^O| (a^O_k)_{ig}}=a \hat{\sigma}_{ig}^k(\alpha,\gamma)}{\sum_{(i,g) \in A_k^O}  \hat{\sigma}_{ig}^k(\alpha,\gamma)}
\end{equation}

\section{Expectation-maximization algorithm}

To obtain a maximum of the posterior we start by berating random initial conditions for each model parameter $\boldsymbol{\theta}, \boldsymbol{\eta}, \boldsymbol{p}, \boldsymbol{\zeta} ,\boldsymbol{q}, \hatqs$.

The we perform iteratively two steps until model parameters convergence:
\begin{enumerate}
    \item Expectation step: compute the auxiliary functions $\omega_{ij}(\alpha,\beta)$, $\sigma^k_{i\ell_k}(\alpha)$, and  $\hat{\sigma}^k_{ig}(\alpha,\gamma)$  using current values for  $\boldsymbol{\theta}, \boldsymbol{\eta}, \boldsymbol{p}, \boldsymbol{\zeta} ,\boldsymbol{q}, \hatqs$ using Eqs. A.4, A.5 and A.6.
    \item Maximization step: Compute the new values for the model parameters using the values for the auxiliary functions and Eqs. A.7 - A.12.
\end{enumerate}


Because the posterior landscape is very rugged, to make predictions we perform the EM algorithm 10 times and consider all of the models to estimate the average probability that  user $i$ rates  item $j$ with rating $r$ (see \cite{godoy-lorite16b}) as follows:
\begin{equation}
    \langle p(r_{ij}=r|R^O, A^O_k)\rangle \approx \frac{1}{N}\sum_{n=1}^N p_n(r_{ij}=r|R^O, A^O_k,(\dots))
\end{equation}
where $(\dots)=\{ \boldsymbol{\theta}, \boldsymbol{\eta}, \boldsymbol{p}, \boldsymbol{\zeta} ,\boldsymbol{q}, \hatqs\}$, and $p_n(r_{ij}=r|R^O, A^O_k,(\dots))$ is the probability  that  user $i$ rates  item $j$ with rating $r$ in run $n$ of the EM algorithm.

\bibliography{ref-database,local-refs}

\end{document}